# Superconducting and Normal State Properties of Heavily Hole-Doped Diamond

V. A. Sidorov[1,2], E. A. Ekimov[1], S. M. Stishov[1,2], E. D. Bauer[2], and J. D. Thompson[2]

[1]*Vereshchagin Institute for High Pressure Physics, Russian Academy of Sciences, 142190 Troisk, Moscow Region, Russia*

[2]*Los Alamos National Laboratory, Los Alamos, NM 87545*

Abstract: We report measurements of the specific heat, Hall effect, upper critical field and resistivity on bulk, B-doped diamond prepared by reacting amorphous B and graphite under high-pressure/high-temperature conditions. These experiments establish unambiguous evidence for bulk superconductivity and provide a consistent set of materials parameters that favor a conventional, weak coupling electron-phonon interpretation of the superconducting mechanism at high hole doping.

The theoretical prediction [1] and nearly simultaneous discovery of superconductivity in self-doped degenerate semiconductors, such as $Ge_xTe$ [2], $Sn_xTe$ [3] and $SrTiO_{3-\delta}$ [4], was, at the time, an important validation of the recently developed BCS theory of superconductivity and of current understanding of electron-electron and electron-phonon interactions. The basic premise of these predictions was that superconductivity would be favored in doped semiconductors with many-valley band structures due to their relatively enhanced density of electronic states and the attractive interaction provided by inter-valley phonon scattering. As expected, the superconducting transition temperatures of these systems were relatively low, with $T_c$s < 0.5 K. Diamond-structured Si and Ge, which have mutli-valley band structures [5], likewise were predicted to have $T_c$s near 0.005 K when carrier doped at a concentration $\approx 10^{20}$ cm$^{-3}$ [1], but superconductivity has not been found in their diamond structure. Recently, however, superconductivity was reported in diamond itself when hole-doped by B additions. [6] Because of its small atomic radius, B is incorporated relatively easily into the dense diamond lattice and, with one less electron than C, dopes holes into a shallow acceptor level close to the top of the valence band. Assuming that each of the approximately $\approx 4.9 \times 10^{21}$ B/cm$^3$ in this superconductor contributed 1 hole/B to diamond, this hole density exceeded $n_{IM} \sim 2 \times 10^{20}$ cm$^{-3}$ that is necessary to induce an insulator-metal transition; beyond this carrier concentration, the impurity band formed by donor states begins to overlap the valence band edge.[7] The superconducting transition temperature, $T_c \approx 4$K, of hole-doped diamond is substantially higher than predicted for diamond-structured Si or Ge, even if they were doped hypothetically to n $\sim 10^{21}$cm$^{-3}$. [8]

Besides the existence of superconductivity, relatively little else is known to constrain interpretations of the superconducting mechanism in diamond. On the basis of existing information, two qualitatively different views have emerged: superconductivity is electron-phonon mediated [9-12] or arises from a resonating valence bond type of mechanism [13]. Though differing in detail, models of a conventional mechanism conclude that holes doped into the degenerate σ-bonding valence band are coupled most strongly by zone-center optical phonon modes that soften as holes are added. This mechanism is analogous to that producing superconductivity near 40 K in $MgB_2$, [14] but because of the 2-dimensional character of graphitic B layers in $MgB_2$, holes are much more strongly coupled to optical phonon modes and $T_c$ correspondingly is an order of magnitude higher than in diamond. With conventional values for the Coulomb pseudopotential μ* =0.1 to 0.2, these electron-phonon models for superconducting diamond give correct estimates of $T_c$, within factors of order unity. However, the small carrier concentration in superconducting diamond implies that the Coulomb interaction may be poorly screened, opening the possibility of an exotic pairing mechanism. This different viewpoint [13] rests, in part, on the premise that the carrier doping level in superconducting diamond is close to the Mott limit. In this case, the disordered lattice of B and associated random Coulomb potential lift the orbital degeneracy of B acceptor states, producing a single, narrow half-filled band in which superconductivity could arise through a resonating valence bond (RVB) type mechanism. Here we report the first specific heat measurements on newly prepared superconducting diamond that, with Hall effect, resistivity, upper critical field, and magnetic susceptibility measurements, clearly establish the bulk nature of superconductivity and appear to rule out an exotic pairing mechanism.

In the initial report of superconducting diamond [6], B doping was achieved by reacting graphite and $B_4C$ at high pressures (8-9 GPa) and temperatures (2500-2800 K). Under these conditions, small B-doped diamond aggregates formed that were sufficiently large for electrical transport and magnetic susceptibility measurements. These experiments showed the onset of a resistive transition to the superconducting state near 4 K and zero-resistance below 2.3 K where a strong diamagnetic response developed. Though these measurements provided strong evidence for bulk superconductivity, the small sample mass and small heat capacity of B-doped diamond prevented the detection of a clear specific heat anomaly at $T_c$ that would confirm the bulk nature of superconductivity. We have synthesized more massive aggregates of bulk B-doped diamond using similar high-pressure, high-temperature conditions but starting with a mixture of amorphous B powder and graphite in which the B content was 4 wt.%. Details of sample-preparation procedures and structural characterization will appear elsewhere. [15] Analysis of x-ray diffraction patterns on the resulting diamond aggregates gives a cubic lattice parameter of 3.573 Å, which is larger than the lattice constant of pure diamond (3.5664 Å) and indicates that B is incorporated into the diamond lattice. Hall-effect measurements at room temperature on this new sample give, within a single band approximation, a hole concentration of $n_H$~$1.8 \times 10^{21}$ cm$^{-3}$. Earlier studies have shown that the Hall number in B-doped diamond is nearly temperature independent with decreasing temperature. [16] For comparison, the concentration of dissolved boron in the diamond lattice, estimated from Vegard's law and the measured lattice parameter [17], is ~$2.5 \times 10^{21}$ B/cm$^3$, which agrees reasonably with the value derived from Hall-effect measurements. The hole and deduced B concentrations are both about half that *inferred* from the B concentration ≈ $4.9 \times 10^{21}$ B/cm$^3$ in earlier samples prepared from $B_4C$ and graphite. Comparable or larger discrepancies of the same sign between B concentration and Hall number are found [16] in B-doped diamond films, but the origin of these discrepancies is poorly understood. Because doped holes go predominantly into a single, σ-bonding valence band,[9-12] a single band interpretation of Hall-effect measurements should reflect the carrier concentration rather accurately. The B concentration, on the other hand, could misrepresent the hole density due to the presence of compensating n-type impurities, such as nitrogen, or if not all B atoms were incorporated into the diamond lattice. In the following, we use the Hall number, where it is available, as the relevant measure of the carrier concentration.

Figure 1a shows the temperature dependence of the electrical resistivity $\rho(T)$ of the new sample prepared with B powder and measured using a standard 4-probe technique. The overall shape of the curve, including a metallic-like resistivity above about 250 K, is very similar to that reported earlier [6]. The room-temperature resistivity of the present sample agrees quantitatively with reports on B-doped diamond films having the same hole concentration determined by Hall-effect measurements. [16, 18] The inset of Fig. 1a shows the onset of a transition near 4.5 K and an immeasurably small resistance below 3.4 K. Both temperatures are comparable to but somewhat higher than those reported earlier.[6] This broadened resistive transition indicates an inhomogeneous distribution of charge carriers that is reflected as well in *ac* susceptibility $\chi_{ac}$ measurements given in Fig. 1b. In contrast to the very sharp transition at 7.2 K in a comparably sized piece of Pb in the measurement coil, $\chi_{ac}$ of B-doped diamond starts to deviate from its temperature-independent value near 4.5 K and is followed at ~2.3 K by a much steeper transition to a fully diamagnetic state. These resistance and *ac* susceptibility measurements establish that superconductivity in B-doped diamond is a robust result that does not depend quantitatively on starting materials.

Application of a magnetic field suppresses the resistive transition, as shown in the inset of Fig. 2. With increasing field, $\rho(T)$ broadens, with the zero-resistivity temperature decreasing somewhat more rapidly than the onset temperature. Upper critical fields $H_{c2}(T)$, defined by the temperatures at which $\rho(T)$ reaches 10% and 90% of its value just above the resistive onset, are plotted in the main panel of Fig. 2. The resulting slopes $\partial H_{c2}/\partial T$ are more than twice the slope $\partial H_{c2}/\partial T = -1.7$ T/K reported earlier.[6] An extrapolation of these data to T=0 using the standard dirty-limit expression for a type-II superconductor $H_{c2}(0)=-0.69(\partial H_{c2}/\partial T)T_c$ [19] gives $H_{c2}(0) \approx 14$ T for the higher transition and ≈9.0 T for the lower transition and corresponding Ginzburg-Landau coherence lengths $\xi_{GL}=[\Phi_0/2\pi H_{c2}(0)]^{1/2} \approx 48$ and 60 Å, respectively.

Specific heat was measured on a 43-mg sample taken from the same growth product that also gave samples used for resistivity and susceptibility measurements. As seen in Fig. 3, there are anomalies in specific heat divided by temperature C/T that correspond closely in temperature to those observed in $\chi_{ac}$. Because specific heat is a thermodynamic measure of bulk properties, the existence of these anomalies establishes unambiguously that superconductivity in B-doped diamond develops throughout the sample volume. A fit of these data above 4.5 K to $C/T = \gamma+\beta T^2$ gives an average electronic contribution $\gamma = 0.113$ mJ/mol K$^2$ and, from the value of β, a sample-average Debye temperature $\Theta_D = 1440$ K, which is about 23% smaller than the Debye temperature of undoped diamond and indicates lattice softening in response to doped holes. Assuming an ideal, uniformly B-doped sample with a $T_c$ of 4 K, we construct a curve shown in Fig. 3 that conserves measured entropy in the superconducting and normal states. This idealized curve provides a rough estimate of the specific heat jump at this $T_c$, $\Delta C/\gamma T_c = 0.5$, which is about one-third of the value 1.43 expected for weak coupling superconductivity. Very similar values of $\Delta C/\gamma T_c$ also are found in superconducting $Ge_{0.950}Te$ [20] and $Sn_{0.975}Te$ [21].

Data shown in Figs.1-3 imply some form of inhomogeneity in the present sample that also was the case in a sample made with $B_4C$ [6]. There are several potential sources of non-uniformity in B-doped diamond. Boron plays a double role in the synthesis route we use to produce superconducting diamond: it serves as a catalyst, facilitating graphite transformation into diamond, and, at the same time, B is captured by the diamond lattice as a hole dopant. Upon cooling from the high temperature synthesis conditions, some diamond grains may react more efficiently with boron to form an intergrowth structure having different B concentrations. There are two additional factors, possibly related to the first, that contribute to inhomogeneity of B-doped diamond, irrespective of the preparation technique: (a) B atoms substitute for C at low doping levels; whereas, at high doping levels, additional boron atoms enter the diamond structure interstitially and strain the lattice [22]; and (b) boron atoms are incorporated preferentially in certain growth sectors [23, 24]. In the absence of specific knowledge of the growth morphology and B distribution in our samples, it is not possible to make definitive statements about the origin of inhomogeneity; although, preferential B uptake in certain growth sectors provides a plausible mechanism to account for the specific heat behaviors of the present sample. [25]

In spite of some inhomogeneity, a simple, consistent interpretation of measured properties is possible. From free-electron relations for the Fermi momentum $k_F=(3\pi^2 n)^{1/3}$ and Sommerfeld coefficient $\gamma_0=k_B^2 m_b k_F/(3\hbar^2)$, we obtain $\gamma_0= (m_b/m_e)68$ $\mu$J/mole-C $K^2$, where $m_b$ ($m_e$) is the band (electron) mass and carrier concentration $n = n_H = 1.8 \times 10^{21}$ $cm^{-3}$ established by Hall measurements. This free-electron estimate for $\gamma_0$ is equal to the sample-average value determined by specific heat if $m_b = 1.7 m_e$, which agrees well with the effective band mass, $\sim 1.3 m_e$, of diamond. [26] The slope of the upper critical field $\partial H_{c2}/\partial T$ provides another consistency check. In the limit that the electronic mean free path $l_{mfp}$ is less than $\xi_{GL}$, $\partial H_{c2}/\partial T = -A\rho(T_c)\gamma$, where A is a units-dependent constant. Taking the measured values of $\rho(T_c)= 2.5$ m$\Omega$cm and $\gamma = 0.113$ mJ/mol $K^2$ gives $\partial H_{c2}/\partial T = -3.7$ T/K, which is within 10% of the value determined using the $0.1\rho(T \geq T_c)$ criterion to define $H_{c2}(T)$ shown in Fig. 2. Justification for the dirty-limit assumption is provided from a free-electron estimate of $l_{mfp} = \hbar k_F/(\rho n e^2) \approx 3.4$ Å $<< \xi_{GL}$. These comparisons neglect inhomogeneity in the carrier concentration. In spite of this, inhomogeneity-induced variations of approximately ±25%, and not factors of two or more, in the parameters would account for experimental observations, including two superconducting transitions separated in temperature by ~2K. Though providing a reasonable starting point for comparison to model calculations, this level of uncertainty precludes a meaningful estimate of the electron-phonon coupling parameter $\lambda$ that has been predicted

[9-12] to range from ≈0.3 to ≈0.5 in diamond doped with B at concentrations comparable to or slightly higher than in the present experiments. In principle, $\lambda$ could be estimated from $\gamma_{meas} = \gamma_0(1+\lambda)$. Stated alternatively, the band mass in free-electron relations used above should be replaced by $m^* = m_b(1+\lambda)$. Part of the discrepancy between the estimated $m_b$ and average band mass of diamond could be due to electron-phonon coupling corrections, but it is impossible to draw this conclusion at present. Nevertheless, our results imply that $\lambda$ for B-doped diamond is much smaller than $\lambda \approx 1$ in $MgB_2$, as also suggested in Refs. 9 and 10. Taken together, the general consistency of free-electron like parameters derived from these experiments argue against the importance of electronic correlations and an exotic pairing mechanism for superconductivity in B-doped diamond, at least at these relatively high doping levels.

Because of potential technological applications, B-doped diamond films have been studied more extensively than bulk diamond prepared under high pressure/high temperature conditions. [27] Such films, in principle, are amenable to systematic studies of the carrier dependence of $T_c$ that would further constrain proposed mechanisms of superconductivity. [28] We [29] as well as others [28, 30] have observed superconductivity in chemically vapor deposited (CVD) films of B-doped diamond. Preliminary studies are encouraging but also emphasize the extent to which superconducting properties depend on film-growth conditions. For example, a B-doped film grown with predominantly {111} facets and having a Hall number $n_H = 0.9 \times 10^{21}$ cm$^{-3}$ exhibits a resistive onset near 7 K and zero resistance below 4.2 K. [30] Depending on the criterion used to define $H_{c2}(T)$, the extrapolated $H_{c2}(0)$ ranges from 5 to 10 T. On the other hand, a homoepitaxial film with a boron concentration of $1.9 \times 10^{21}$ B/cm$^3$ and growing with a {001} facet has a relatively sharp zero-field resistive transition near 2 K that broadens substantially in magnetic fields below $H_{c2}(0) \approx 1.5$T. [28] For this particular growth habit, superconductivity disappears as the B concentration approaches the insulator-metal boundary. [29] To the extent that a comparison is possible, trends in superconductivity and inhomogeneity found in these CVD films are qualitatively similar to those in bulk, B-doped diamond and suggest an avenue of investigation that should be pursued to clarify quantitative responses of both films and bulk samples of B-doped diamond.

In summary, specific heat, Hall effect, upper critical field and resistivity measurements on bulk, B-doped diamond establish unambiguous evidence for bulk superconductivity and provide a consistent set of materials parameters that favor a conventional, weak coupling electron-phonon mechanism in these heavily hole-doped samples. This conclusion is consistent with theoretical models [9-12] that draw an analogy between hole-doped diamond and more nearly 2-dimensional $MgB_2$. Though an exotic pairing mechanism may become operable as hole doping is reduced toward the insulator-metal boundary, preliminary studies of B-doped thin films find a monotonic decrease of $T_c$ with decreasing B concentration, [28] suggesting a common pairing mechanism at low and high B concentrations. Because of the inhomogeneous incorporation of B in diamond, systematic studies, including Hall measurements, on mono-faceted, fully dense samples would help refine more precisely materials parameters for a quantitative comparison to model calculations and for an inter-comparison of experimental results.

Acknowledgements: V. A. S., E. A. K., and S. M. S. acknowledge support from the Russian Foundation for Basic Research Grant No. 03-02-17119 and Programs "Strongly Correlated Electrons" and "Physics of High Pressures" of the Department of Physical Sciences, Russian Academy of Sciences. Work at Los Alamos was performed under the auspices of the U.S. DOE/Office of Science.

impurity band may not be able to support superconductivity but still contribute a finite density of electronic states that is suggested from the extrapolated non-zero value of C/T as T→0. How these sources of real-space inhomogeneity might affect electronic structure and their possible relevance to our observations are unknown but deserve theoretical and experimental attention.

Figure Captions:

Fig.1. (a) Resistivity as a function of temperature for bulk, B-doped diamond. The inset shows the onset of a superconducting transition near 4.5 K below which the resistivity becomes immeasurably small at 3.4 K. (b) *AC* susceptibility of the B-doped diamond and a comparably sized piece of Pb in a counter-wound coil. The inset shows the onset of a diamagnetic response in B-doped diamond at 4 K that is followed below 2.3 K by a response comparable to that of Pb with $T_c$=7.22 K.

Fig. 2. Upper critical magnetic field as a function of temperature for bulk, B-doped diamond. Curves were constructed by taking the field-dependent temperatures at which ρ(T), plotted in the inset, reached 90% (open circles) and 10% (filled circles) of its value just above the resistive onset. Representative data in the inset are, from right to left, for fields of 0, 1, 3, 5, 7 and 9 T.

Fig. 3. Specific heat divided by temperature versus temperature for bulk, B-doped diamond. Two anomalies, corresponding closely in temperature to those found in *ac* susceptibility, confirm the bulk nature of superconductivity. The dashed line is an entropy-conserving construction, assuming an idealized transition at 4 K. A linear fit of C/T versus $T^2$ for T ≥ 4.5 K, plotted in the inset, gives a sample-averaged electronic Sommerfeld coefficient γ = 0.113 mJ/mol-C $K^2$ and $\Theta_D$ = 1440 K.

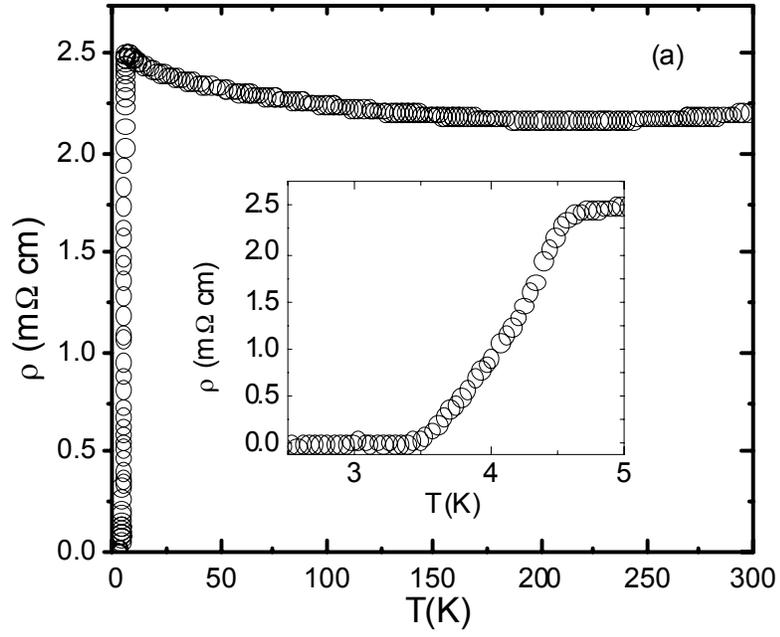
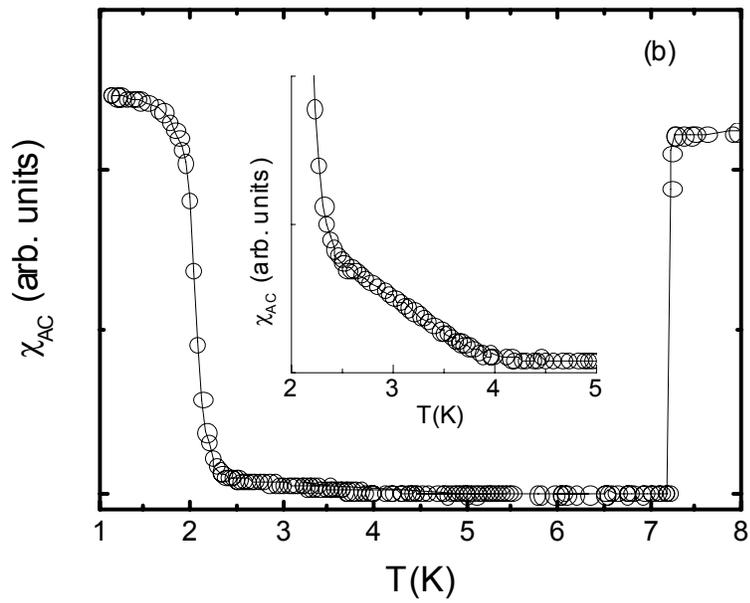

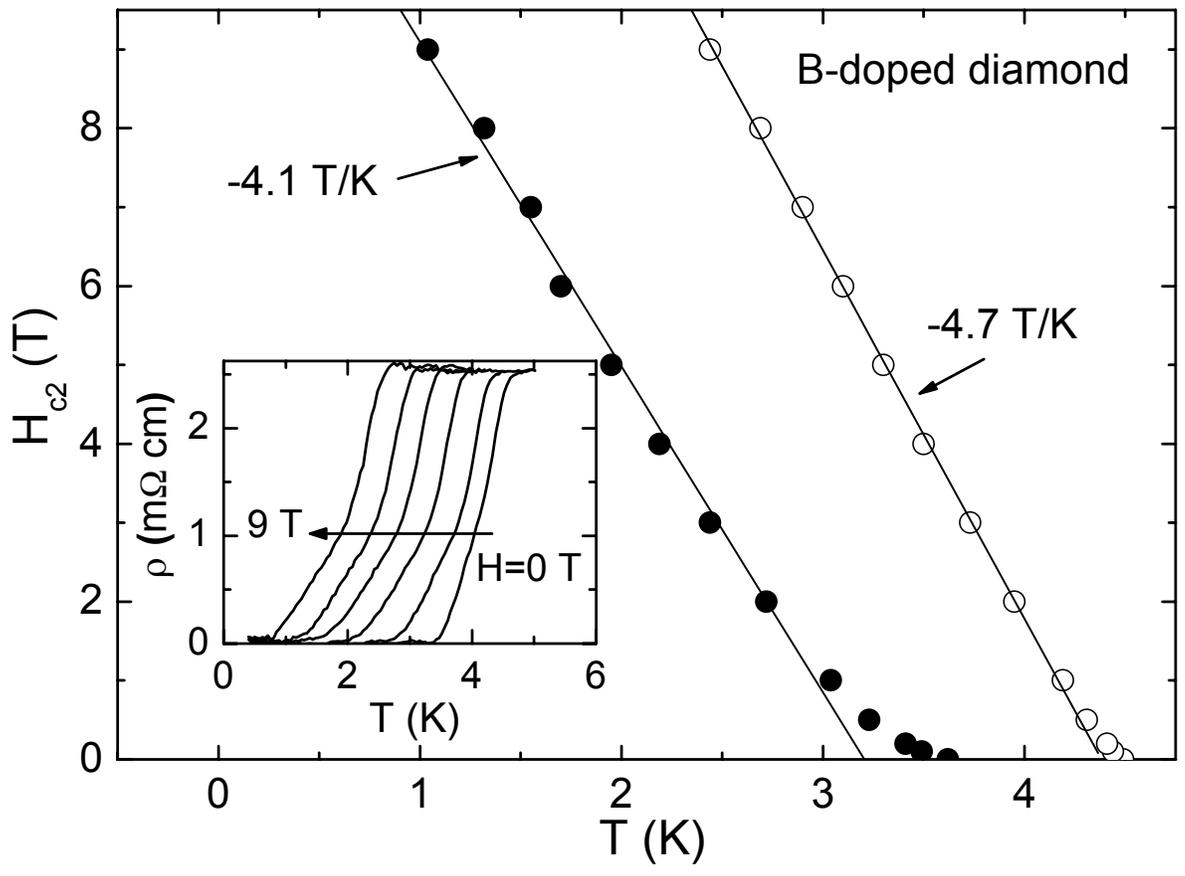

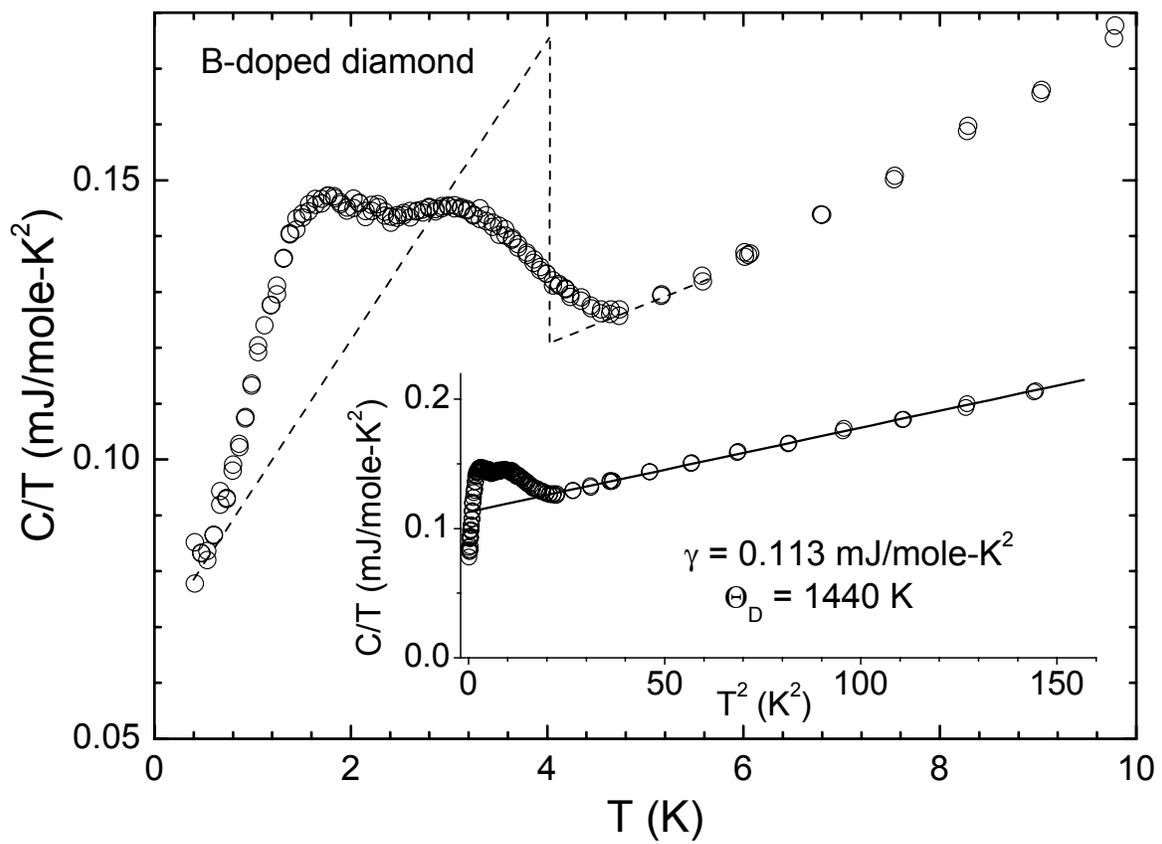